\documentclass[aps,prd,floatfix,superscriptaddress,preprintnumbers,amsmath,amssymb]{revtex4}
\usepackage{amssymb}
\usepackage{amsmath}
\usepackage{amsfonts}
\usepackage{epsfig}
\usepackage{graphicx}
\usepackage{tabularx}
\usepackage{color}

\newcommand{\seq}{\begin{subequations}}
\newcommand{\sen}{\end{subequations}}
\newcommand{\eq}{\begin{eqnarray}}
\newcommand{\en}{\end{eqnarray}}

\def\shiftdown#1{#1\llap{\lower.04ex\hbox{#1}}}

\newcommand{\bfq}{{\bf q}_{\perp}}

\newcommand{\bfk}{{\bf k}_{\perp}}

\begin{document}

\title{Baryons in a soft-wall AdS-Schwarzschild approach at low temperature} 

\author{Thomas Gutsche}
\affiliation{Institut f\"ur Theoretische Physik,
Universit\"at T\"ubingen, \\
Kepler Center for Astro and Particle Physics,
Auf der Morgenstelle 14, D-72076 T\"ubingen, Germany}
\author{Valery E. Lyubovitskij}
\affiliation{Institut f\"ur Theoretische Physik,
Universit\"at T\"ubingen, \\
Kepler Center for Astro and Particle Physics,
Auf der Morgenstelle 14, D-72076 T\"ubingen, Germany}
\affiliation{Departamento de F\'\i sica y Centro Cient\'\i fico
Tecnol\'ogico de Valpara\'\i so-CCTVal, Universidad T\'ecnica
Federico Santa Mar\'\i a, Casilla 110-V, Valpara\'\i so, Chile}
\affiliation{Department of Physics, Tomsk State University,
634050 Tomsk, Russia}
\author{Ivan Schmidt}
\affiliation{Departamento de F\'\i sica y Centro Cient\'\i fico
Tecnol\'ogico de Valpara\'\i so-CCTVal, Universidad T\'ecnica
Federico Santa Mar\'\i a, Casilla 110-V, Valpara\'\i so, Chile}
\author{Andrey Yu. Trifonov}
\affiliation{Laboratory of Particle Physics, 
Tomsk Polytechnic University, 634050 Tomsk, Russia} 

\vspace*{.2cm}

\date{\today}

\begin{abstract}

Recently we derived a soft-wall AdS-Schwarzschild approach 
at small temperatures for the description of hadrons with integer 
spin and adjustable number of constituents (mesons,  tetraquarks, 
dibaryons, etc.). In the present paper we extend our formalism to 
states with half-integer spin (baryons, pentaquarks, etc.), 
presenting analytical results for the temperature dependence of 
their masses and form factors.   

\end{abstract}

\maketitle

\section{Introduction}

The study of the temperature dependence of hadron properties gives 
an opportunity for getting a deeper understanding of several physical 
phenomena, such as the evolution of the early Universe, and 
the formation and phase transitions in both hadronic and nuclear matter.  
One of the powerful methods to get insights into the 
thermal properties of hadrons is holographic QCD.  
For the recent progress achieved by holographic QCD in this direction 
see Refs.~\cite{Herzog:2006ra}-\cite{Gutsche:2019blp}, and 
Ref.~\cite{Gutsche:2019blp} for a short overview. 
In Ref.~\cite{Gutsche:2019blp} we proposed a modification of 
the soft-wall model at finite temperature $T$ in order to have 
consistency with QCD properties. In particular, we argue that in order to 
reproduce a temperature behavior of the quark condensate one should include 
a $T$-dependence of the dilaton field (which is the parameter of 
spontaneous breaking of chiral symmetry related to the pseudoscalar meson 
decay constant) and the warping of the anti–de Sitter (AdS) metric due to 
temperature. In particular, we proposed that the dilaton field has a specific 
$T$-dependence, which is dictated by the temperature behavior of the 
chiral quark condensate in QCD~\cite{Gasser:1986vb}-\cite{Toublan:1997rr}, 
derived using chiral perturbation theory (ChPT)~\cite{ChPT}.
In this way we postulated the temperature dependence of the 
dilaton field, using its relation to the chiral quark condensate at 
zero temperature. A thermal behavior of the dilaton 
has been previously proposed in Ref.~\cite{Vega:2018dgk}. 
In the present case we aim for consistency with QCD, which makes the study 
important in order to improve the understanding of hadronic properties 
at finite temperature. In Ref.~\cite{Gutsche:2019blp} we were 
interested in the low temperature limit and in the derivation of analytical 
formulas for the mass spectrum of mesons and their form factors within 
a soft-wall AdS/QCD model~\cite{Karch:2006pv}-\cite{Gutsche:2017oro}. 
In particular, we considered two possible sources of temperature dependence: 
(1) the warping of the AdS metric due to temperature, 
(2) the temperature dependence of the dilaton-background field.  
This field produces confinement and is responsible for the breaking of 
conformal invariance and the spontaneous breaking of chiral symmetry 
in holographic QCD. 

In the present paper we extend the ideas and formalism of 
Ref.~\cite{Gutsche:2019blp} to the case of hadrons with half-integer spin, 
and it is structured as follows.
In Sec.~II we present the details for the  construction of an effective
action for AdS fields with half-integer spin at small temperatures, and 
apply it to the calculation of the mass spectrum and form factors. 
In Sec.~III we discuss a derivation of a new quantity --- 
the hadron light-front wave function at finite temperature. It is obtained 
using the matching of form factors obtained in our approach and 
the Drell-Yan-West (DYW) formula~\cite{Drell:1969km} for 
the hadronic form factors in light-front QCD. 
Finally, in Sec.~IV, we summarize the results of the paper.

\section{Framework} 

\subsection{Effective action and hadron masses at low temperatures}

In this section we start with the derivation of the 
five dimensional action for the fermion bulk field 
$B_{M_1\ldots M_J}$, with arbitrary total half-integer spin $J$ 
at low temperature $T$. Our formalism is based on the 
analogous action at zero temperature~\cite{Gutsche:2011vb}, 
and includes the issues proposed in Ref.~\cite{Gutsche:2019blp}. 
The AdS-Schwarzschild metric is specified by
\eq
ds^2 = e^{2 A(z)} \,
\biggl[ f_T(z) dt^2 - (d\vec{x})^2 - \frac{dz^2}{f_T(z)} \biggr]
\en
where $x=(t,\vec{x}\,)$ is the set of Minkowski coordinates,
$z$ is the holographic coordinate, $R$ is the AdS radius and
$A(z) = \log(R/z)$. Here $f_T(z) = 1 - z^4/z_H^4$,
where $z_H$ is the position of the event horizon, which is 
related to the black-hole Hawking temperature by $T = 1/(\pi z_H)$. 
As in the case of boson AdS fields, we also introduce 
the exponential prefactor $\exp[-\varphi(z)]$ in the effective action. 
It contains the background (dilaton) field $\varphi(z) = \kappa^2 z^2$ 
where $\kappa$ is a scale parameter of the order of a few hundred MeV.
This dilaton field breaks conformal invariance, produces confinement
and is responsible for the  spontaneous breaking of chiral symmetry in
holographic QCD. In addition to the dilaton, we introduce in the action the
thermal prefactor
\eq\label{lambda_Tz}
e^{-\lambda_T(z)}, \quad \lambda_T(z) = \alpha \frac{z^2}{z_H^2}
+ \gamma \frac{z^4}{z_H^4} + \xi \frac{\kappa^2 z^6}{z_H^4} \,,
\en
where the dimensionless parameters $\alpha$, $\gamma$, and $\xi$
parametrize the $z^2$, $z^4$, and $z^6$ thermal corrections. 
The parameters $\gamma$ and $\xi$ have been fixed 
in Ref.~\cite{Gutsche:2019blp} to guarantee gauge
invariance and massless ground-state pseudoscalar mesons $(\pi$,
$K$, $\eta$) in the chiral limit, and to suppress 
the six power of the radial dependence in the holographic potential: 
\eq 
\gamma = \frac{J(J-3)+3}{5}\,, \quad 
\xi = \frac{2}{5}\,. 
\en 
The parameter $\alpha$ encodes the contribution of gravity
to the restoration of chiral symmetry at a critical temperature $T_c$, 
and the term $\alpha z^2/z_H^2$ was considered to be a small perturbative 
correction to the quadratic dilaton $\varphi(z)$.  Then, for 
convenience, we related the holographic coordinate $z$ to 
the Regge-Wheeler (RW) tortoise coordinate $r$ 
via the substitution~\cite{Regge:1957td,Horowitz:1999jd}:
\eq\label{RW_variable}
r = \int \frac{dz}{f_T(z)} = \frac{z_H}{2} \,
\biggl[ - {\rm arctan}\frac{z}{z_H}
+ \frac{1}{2} \, \log\frac{1-z/z_H}{1+z/z_H}\biggr] \,.
\en
Here we use the plus sign in the right-hand side (r.h.s.)  
of Eq.~(\ref{RW_variable}). As in Ref.~\cite{Gutsche:2019blp}, 
we restrict ourselves to the leading-order (LO) and 
next-to-leading-order (NLO) terms in the expansion of $z$ in powers of $r$: 
\eq\label{expansion_r}
z = r \biggl[ 1 - \frac{t_r^4}{5} 
+ {\cal O}\Big(t_r^{8}\Big)  \biggr]\,, \quad t_r = r/z_H \,.
\en
Using~(\ref{expansion_r}) the generalized 
exponential prefactor in the case of the boson AdS bulk fields reads  
\eq\label{calP}
{\cal P} = \exp\Big[-\varphi_T(r) - \gamma \frac{r^4}{z_H^4}\Big] \,. 
\en 
In the following we construct the action for fermion AdS fields 
$B_{M_1\ldots M_J}$ dual to hadrons with total half-integer 
spin $J$ and number of the constituents $N$.  
As was shown in Ref.~\cite{Gutsche:2011vb} 
in the case of baryons the exponential prefactor in the action 
can be absorbed by a redefinition of AdS fermion fields. 
Here we extend this idea to the prefactor ${\cal P}$,
performing redefinition the AdS fermion field ${\cal B}_{M_1\ldots M_J}$ 
as 
\eq 
{\cal B}_{M_1\ldots M_J}(x,r,T) =  
{\cal P}^{-1/2} \ B_{M_1\ldots M_J}(x,r,T) \,. 
\en 
Now the action for fermion 
bulk fields $B_{N_1\ldots N_J}(x,r,T)$ reads:  
\eq\label{action_SM}
S_B &=& \int d^4x dr \, \sqrt{g} \ 
\bar B_{N_1\ldots N_J}(x,r,T) \, \hat{\cal D}_\pm(r) \,  
B^{N_1\ldots N_J}(x,r,T) \,, \nonumber\\
\hat{\cal D}_\pm(r) &=&  \frac{i}{2} \Gamma^M \Big[
\! \stackrel{\leftrightarrow}{\partial}_{_M} - \frac{1}{4}
\, \omega_M^{ab} \, [\Gamma_a, \Gamma_b] \Big]
\, \mp \,  \Big[ \mu(r,T) + U_F(r,T) \Big]\,. 
\en
Here $\mu(r,T) = \mu/f_T^{3/10}(r)$ is the temperature dependent 
five-dimensional mass of the AdS fermion with half-integer spin  
$\mu = N + L - 3/2$, where $N$ and $L$ are the number of partons 
and orbital angular momentum, respectively. 
$U_F(r,T) = \varphi_T(r)/f_T^{3/10}(r)$ is the dilaton temperature 
dependent potential, in which $\varphi_T(r)$ is the $T$-dependent 
dilaton field derived in Ref.~\cite{Gutsche:2019blp}: 
\eq\label{dilaton_fT}
\varphi_T(r) &=& K_T^2 r^2 = (1 + \rho_T) \kappa^2 r^2 \,, 
\nonumber\\
\rho_T &=& \biggl(\frac{9 \alpha \pi^2}{16}
\,+\, \delta_{T_1}\biggr) \frac{T^2}{12 F^2}
\,+\, \delta_{T_2} \biggl(\frac{T^2}{12 F^2}\biggr)^2
\,+\,{\cal O}(T^6) \,. 
\en 
The quantity $\rho_T$ encodes the $T$-dependence of 
the dilaton, in the form of an $T^2/(12 F^2)$ expansion 
dictated by QCD~\cite{Gasser:1986vb} and $F$ is 
the pseudoscalar coupling constant in the chiral limit;
$\omega_M^{ab} = (\delta^a_M \delta^b_r 
               - \delta^b_M \delta^a_r)/(r f_T^{1/5}(r))$ is
the spin connection term; $\sigma^{MN} = [\Gamma^M, \Gamma^N]$
is the commutator of the Dirac matrices in AdS space, which are defined as
$\Gamma^M = \epsilon^M_a \Gamma^a$ and
$\Gamma^a = (\gamma^\mu, -i \gamma^5)$. The term with partial derivative 
is defined as 
\eq
\Gamma^M \! \stackrel{\leftrightarrow}{\partial}_{_M}
= \Gamma^M \! \Big(\stackrel{\leftarrow}{\partial}_{_M} -
\stackrel{\rightarrow}{\partial}_{_M}\Big)
= g^{MN} \epsilon^a_N \Gamma_a \Big(\stackrel{\leftarrow}{\partial}_{_M} -
\stackrel{\rightarrow}{\partial}_{_M}\Big)\,, \quad 
\epsilon^a_N = \frac{R}{z} \, \delta^a_N \,. 
\en 

Using the axial gauge $B_z(x,r,T) = 0$ we expand the 
fermion AdS field into left- and right-chirality 
components: 
\eq 
& &B_{N_1\ldots N_J}(x,r,T) = B^R_{\mu_1\ldots \mu_J}(x,r,T) 
\, + \, B^L_{\mu_1\ldots \mu_J}(x,r,T) \,, \nonumber\\ 
& &B^{L/R} = \frac{1 \mp \gamma^5}{2} \, B \,, \quad 
\gamma^5 B^{L/R} = \mp B^{L/R} \,. 
\en 
Then we perform a Kaluza-Klein expansion for the four-dimensional 
transverse components of the AdS fields 
\eq\label{KK_coord}
B^{L/R}_{\mu_1\ldots \mu_J}(x,r,T) = 
\sum\limits_n \ B^{L/R}_{n, \mu_1\ldots \mu_J}(x) \, 
\Phi^{L/R}_{nJ}(r,T), 
\en
where $n$ is the radial quantum number and $B^n_{\mu_1\ldots \mu_J, n}(x)$
is the tower of the Kaluza-Klein (KK) modes, dual to hadrons with 
half-integer spin $J$.  
$\Phi^{L/R}_{nJ}(r,T)$ are their extradimensional profiles (wave functions) 
depending on the temperature. 
After substituting 
$\Phi^{L/R}_{nJ}(r,T) = e^{A(r) (J-2)} \, \phi^{L/R}_{nJ}(r,T)$ 
in the rest frame of the AdS field with $\vec{p} = 0$, 
one can derive the Schr\"odinger-type equations of 
motion~\cite{Gutsche:2011vb} for $\phi^{L/R}_{nJ}(r,T)$
\eq\label{Shroedinger_Eq}
\biggl[ -\partial_z^2 + U_{L/R}(r,T) \biggr] \phi^{L/R}_{nJ}(r,T) = 
M_n^2(T) \, \phi^{L/R}_{nJ}(r,T) \,. 
\en
$U_{L/R}(r,T)$ is the effective potential 
at finite temperature for the
left/right bulk profile $\phi^{L/R}_{nJ}(r,T)$. 
It can be decomposed into a zero temperature term 
$U_{L/R}(r) \equiv U_{L/R}(r,0)$ and a temperature dependent term
$\Delta U_{L/R}(r,T)$ 
\eq 
U_{L/R}(r,T) &=& U_{L/R}(r) \,+\, \Delta U_{L/R}(r,T)
\,, \nonumber\\
U_{L/R}(r) &=& \kappa^4 r^2 \, + \, 
2 \kappa^2 \Big( m \mp \frac{1}{2} \Big) 
\, + \, \frac{m (m \pm 1)}{r^2} \,, 
\nonumber\\
\Delta U_{L/R}(r,T) &=& 2 \rho_T \kappa^2 \, 
\Big(\kappa^2 r^2 + m \mp \frac{1}{2} \Big), 
\en 
where $m = N + L - 3/2 = \tau - 3/2$. 

Note that Eq.~(\ref{Shroedinger_Eq}) is solved using 
the boundary conditions for the modes $\phi^{L/R}_{mJ}(r,T)$ 
in the ultraviolet (UV) and infrared (IR) limits: 
\eq
\phi^{L/R}_{nJ}(r,T) \ \sim \ r^{N+L-1 \pm 1/2}
\ \ {\rm at \ small} \ r\,, \quad\quad
\phi^{L/R}_{nJ}(r,T) \to 0 \ \ {\rm at \ large} \ r\,.
\en
Also the normalizable modes $\Phi_{nJ}(r,T)$ and $\phi_{nJ}(r,T)$
obey the following normalization conditions:
\eq\label{Norm_Cond}
\int\limits_0^\infty dr \, e^{2A(r) (2 - J)} \, \Phi^{L/R}_{mJ}(r,T) 
\Phi^{L/R}_{nJ}(r,T) =
\int\limits_0^\infty dr \, 
\phi^{L/R}_{mJ}(r,T) \phi^{L/R}_{nJ}(r,T) = \delta_{mn} \,.
\en

At both zero temperature $T = 0$ and finite temperature  
the Schr\"odinger-type equations of motion (EOMs) have analytical solutions. 
At $T = 0$ the wave function 
\eq\label{phi_r0}
& &\phi^{L/R}_{nJ}(r,0) = \sqrt{\frac{2 \Gamma(n+1)}{\Gamma(n+m_{L/R}+1)}} 
\ \kappa^{m_{L/R}+1}
\ r^{m_{L/R} + 1/2} \ e^{-\kappa^2 r^2/2} \ L_n^{m_{L/R}}(\kappa^2r^2)\,,
\nonumber\\ 
& &m_{L/R} = m \pm \frac{1}{2}
\en 
corresponds to the mass spectrum 
\eq\label{mass2_T0}
M^2_{nJ}(0) = 4 \kappa^2 \Big( n + m + \frac{1}{2} \Big) \,. 
\en 
Note that $m_L = \tau - 1$ and $m_R = \tau - 2$.

At finite $T$ the solution reads 
\eq\label{phi_rT}
\phi^{L/R}_{nJ}(r,T) = \sqrt{\frac{2 \Gamma(n+1)}{\Gamma(n+m_{L/R}+1)}} 
\ K_T^{m_{L/R}+1}
\ r^{m_{L/R}+1/2} \ e^{-K_T^2 r^2/2} \ L_n^{m_{L/R}}(K_T^2r^2)\,, 
\en 
which corresponds to the mass spectrum 
\eq\label{mass2_T}
M^2_{nJ}(T) = 4 K^2_T \Big( n + m + \frac{1}{2} \Big) 
= 4 \kappa^2 (1 + \rho_T) \Big( n + m + \frac{1}{2} \Big) \,. 
\en
Here 
\eq
L_n^m(x) = \frac{x^{-m} e^x}{n!}
\, \frac{d^n}{dx^n} \Big( e^{-x} x^{m+n} \Big) 
\en
are the generalized Laguerre polynomials. The above formulas are degenerate 
for any value of half-integer $J$ and are valid for fermionic hadrons 
composed of $N$ constituents, 
angular orbital momentum $L$, and radial quantum number~$n$. 

\subsection{Form factors of fermionic hadrons at low temperatures}

In this section, following our study in Refs.~\cite{Gutsche:2011vb,%
Gutsche:2019blp}, we proceed to results for the form factors of 
hadrons with half-integer spin (baryons, pentaquarks, etc.) 
at low temperature. The corresponding form factors are induced by 
the coupling of AdS fields dual to hadrons with external vector 
AdS fields dual to the electromagnetic field. First, we calculate 
the vector bulk-to-boundary propagator at low temperatures, 
using the universal action derived in Eq.~(\ref{action_SM}). 
The corresponding EOM for the Fourier transform of 
the bulk-to-boundary propagator $V(Q,r,T)$, 
in Euclidean metric $Q^2 = -q^2$, reads: 
\eq\label{EOM_VQrT}
\partial_r \biggl( \frac{e^{-\varphi_T(r)}}{r}  \,
\partial_rV(Q,r,T) \biggr) - Q^2 \frac{e^{-\varphi_T(r)}}{r}  \,
V(Q,r,T) = 0 \,.
\en 
This EOM is solved using the boundary conditions for the mode 
$V(Q,r,T)$ in the ultraviolet (UV) and infrared (IR) limits: 
\eq
V(Q,r,T) \ = \ 1 
\ \ {\rm at} \ r = 0\,, \quad\quad 
V(Q,r,T) \to 0 \ \ {\rm at \ large} \ r\,.
\en
EOM~(\ref{EOM_VQrT}) is similar to the EOM for the case of zero
temperature, and the only difference is that the temperature
dependence is absorbed in the $T$-dependence of the
dilaton parameter. Therefore, the solution for
the bulk-to-boundary propagator at small temperature
is straightforward~\cite{Grigoryan:2007my}:
\eq\label{VQ_smallT}
V(Q,r,T) &=& \Gamma(1 + a_T) \, U(a_T,0,K_T^2 r^2)
= K_T^2 r^2 \int_0^1 \frac{dx}{(1-x)^2}
\, x^{a_T} \, e^{- K_T^2 r^2 \frac{x}{1-x} } \,, \quad 
a_T = \frac{Q^2}{4 K_T^2} \,,
\en
where $\Gamma(n)$ and $U(x,y,z)$ are the gamma and Tricomi
functions, respectively.
Now we can calculate the form factor $F_{nJ}(Q^2,T)$, depending
on the Euclidean momentum squared $Q^2$, for fermionic hadrons
with quantum numbers $(n,J,L)$ and number of constituents $N$ 
at low temperature. The master formula is~\cite{Gutsche:2011vb}
\eq
F^{L/R}_{nJ}(Q^2,T) = \int\limits_0^\infty dr 
\Big(\phi^{L/R}_{nJ}(r,T)\Big)^2 \,
V(Q,r,T) \,.
\en
Note that at finite temperature the form factor $F_{nJ}(Q^2,T)$ is 
properly normalized with $F_{nJ}(0,T) = 1$, because of $V(0,r,T) = 1$ 
and $\int\limits_0^\infty dr \Big(\phi^{L/R}_{nJ}(r,T)\Big)^2 = 1$. 
The form factor $F^{L/R}_{nJ}(Q^2,T)$ has the correct power scaling 
at large $Q^2$ consistent with quark counting rules: it is independent 
of the quantum numbers $n$ and $J$, but depends on the number of 
constituents $N$ and the orbital angular momentum $L$: 
\eq 
F^{L/R}_{nJ}(Q^2) \sim \frac{1}{(Q^2)^{m_{L/R}+1}}
\en 
or 
\eq 
F^{R}_{nJ}(Q^2) \sim \frac{1}{(Q^2)^{\tau-1}}\,, \quad 
F^{L}_{nJ}(Q^2) \sim \frac{1}{(Q^2)^{\tau}}\,.  
\en 

Using Eqs.~(\ref{phi_r0}) and~(\ref{VQ_smallT}) we get for 
the ground state $n=0$ fermion 
\eq 
F^{L/R}_{0J}(Q^2,T) = \frac{\Gamma(a_T+1) \, 
\Gamma(m_{L/R}+2)}{\Gamma(a_T+m_{L/R}+2)}\,.  
\en 
Results for radial excitations with any value for $n$ are readily obtained. 
For example, for the first two radial excitations $n=1$ and $n=2$ 
the form factors are 
\eq
F_{1J}(Q^2,T) &=& \frac{\Gamma(a_T+1) \, 
\Gamma(m_{L/R}+4)}{\Gamma(a_T+m_{L/R}+4)}
\,+\, a_T (m_{L/R}+1) \frac{\Gamma(a_T+2) \, 
\Gamma(m_{L/R}+2)}{\Gamma(a_T+m_{L/R}+4)}
\,,\\
F_{2J}(Q^2,T) &=& 
\frac{\Gamma(a_T+1) \, \Gamma(m_{L/R}+6)}{\Gamma(a_T+m_{L/R}+6)}
\,+\,a_T \frac{\Gamma(a_T+2) \, 
\Gamma(m_{L/R}+3)}{\Gamma(a_T+m_{L/R}+6)} 
\nonumber\\
&\times& \biggl[ (m_{L/R}+5) (2m_{L/R}+3) 
+ \frac{1}{2} (m_{L/R}+1) a_T (a_T+5) 
\biggr] \,. 
\en 
Next we can perform a small $T$-expansion of the form factors. 
For the ground state form factor we get 
\eq 
F_{0J}(Q^2,T) &=& F_{0J}(Q^2,0) + \Delta F_{0J}(Q^2,T)\,, \nonumber\\
F_{0J}(Q^2,0) &=& \frac{\Gamma(a+1) \, 
\Gamma(m_{L/R}+2)}{\Gamma(a+m_{L/R}+2)}\,, 
\nonumber\\
\Delta F_{0J}(Q^2,T) &=& \rho_T a \frac{\Gamma(a+1) \, 
\Gamma(m_{L/R}+2)}{\Gamma(a+m_{L/R}+2)} \, 
\Big[ \psi(a+m_{L/R}+2) - \psi(a+1) \Big] \,,
\en 
where $a = Q^2/(4 \kappa^2)$ and 
$\psi(n) = \Gamma'(n)/\Gamma(n)$ is the polygamma function. 
Note that all results for the fermionic hadron form factors 
follow from the results for mesonic hadrons~\cite{Gutsche:2019blp} 
with the substitution $m \to m_{L/R}$. 

Note that the analytical formulas for masses and form factors 
of hadronic states with half-integer spin and adjustable quantum 
numbers $n$ and $L$ at finite temperature are derived in present 
manuscript for the first time in literature. 
Analysis of these quantities at zero temperature 
started in Ref.~\cite{Abidin:2009hr}, where 
originally the soft-wall AdS/QCD action for the nucleon was proposed. 
It included a term describing the nucleon
confining dynamics and the electromagnetic field,
and their minimal and nonminimal couplings. 
Later, in Ref.~\cite{Vega:2010ns},
this action was used for the calculation of
generalized parton distributions of the nucleon.
In Ref.~\cite{Gutsche:2012bp} it was extended to take into
account higher Fock states in the nucleon and additional couplings
with the electromagnetic field in consistency with QCD constituent
counting rules~\cite{Brodsky:1973kr} for the power scaling of
hadronic form factors at large values of the momentum transfer
squared in the Euclidean region.
In Ref.~\cite{Gutsche:2011vb} soft-wall AdS/QCD was developed
for the description of baryons with adjustable quantum numbers
$n$, $J$, $L$, and $S$. In another development,
in Refs.~\cite{Brodsky:2014yha,Sufian:2016hwn,Chakrabarti:2013dda}, 
the nucleon properties have been analyzed using a Hamiltonian
formalism. However, their calculation of the nucleon electromagnetic
properties ignored the contribution of the non-minimal coupling
to the Dirac form factors, and therefore, the analysis done in
Refs.~\cite{Brodsky:2014yha}-\cite{Chakrabarti:2013dda},
is in our opinion not fully consistent. In Ref.\cite{Sufian:2016hwn}
the ideas of Ref.~\cite{Brodsky:2014yha} have been extended by
the inclusion of higher Fock states in the nucleon, in order to calculate
nucleon electromagnetic form factors in light-front holographic QCD.
In this paper the Pauli form factor is again introduced
by hand, using the overlap of the $L=0$ and $L=1$ nucleon wave function.
Additionally, the expression for the neutron Dirac form factor
has been multiplied by hand by a free parameter $r$. 
Recently, in Ref.~\cite{FolcoCapossoli:2019imm} a version
of the soft-wall AdS/QCD approach with the presence of a modified
warp factor in the metric tensor, was proposed.
Notice that in Ref.~\cite{Gutsche:2011vb} we proved that any modification
of the warp factor in the metric tensor can be compensated by an
appropriate choice of the holographic potential. The form
of such potentials for AdS field with different spins were also
analytically derived in~\cite{Gutsche:2011vb}.

\section{Hadronic light-front wave function at finite temperature}

Here we discuss the derivation of a new quantity --- 
$\psi_M(x,\bfk,T) \equiv  \psi_M(x,\bfk; \mu_0)$, 
the hadron light-front wave function at finite temperature and 
initial scale $\mu_0$, based on the matching of form factors 
obtained in our approach $F_M(Q^2,T)$ 
and the DYW formula~\cite{Drell:1969km} for 
the hadronic form factors in light-front QCD: 
\eq\label{DYW_f}
F_M(Q^2,T) = \int\limits_0^1 dx \, \int\frac{d^2\bfk}{16\pi^3} \,
\psi^\dagger_M(x,\bfk',T) \, \psi_M(x,\bfk,T) \,,
\en 
where $\bfk' = \bfk + (1-x) \bfq$ and $Q^2 = \bfq^2$. 
Here $M$ is the quantum number related to twist $\tau$ 
(number of partons $N$ and orbital angular momentum $L$) as  
\eq 
M=m=N+L-2=\tau-2
\en 
for bosonic hadrons, and 
\eq
M=m_L=m+1/2=N+L-1=\tau-1\,, \quad M=m_R=m-1/2=N+L-2=\tau-2 
\en  
for left- and right-handed fermionic hadrons, respectively. 

For simplicity we consider the case of ground state mesons and baryons 
with $n=0$. 
As a result of the matching 
we derive the following unified expression for the LFWF for bosonic 
and fermionic hadrons at the initial scale $\mu_0$
\eq
\psi_M(x,\bfk,T) = N_M \, \frac{4\pi}{K_T} \,
\sqrt{\log(1/x)} \,  
(1-x)^{\frac{M-2}{2}} \, \exp\biggl[- \frac{\bfk^2}{2K_T^2}
\, \frac{\log(1/x)}{(1-x)^2} \biggr]
\en
where
\eq
N_M = \sqrt{M+1} \,.
\en
Note that the dilaton $K_T$ and function $\psi_M(x,\bfk,T)$ vanish at 
the critical temperature $T_c$~\cite{Gutsche:2019blp}: 
\eq 
\frac{T_c^2}{12 F^2}
= N_f \, \biggl[
\sqrt{\frac{N_f^2+1}{N_f^2-1} - 2 \beta + \beta^2} - 1 
+ \beta \biggr] \,, 
\en
where
\eq
\beta = \frac{9 \alpha \pi^2}{16} \, \frac{N_f}{N_f^2-1} 
\en
and $N_f$ is the number of quark flavors. 

Note the hadron light-front wave function is the basic block for the 
definition of matrix elements and hadronic properties in the light-front QCD.
Using $T$-dependent light-front wave function one can calculate
different parton distributions in hadrons, form factors, and structure 
functions  (see for details, e.g., Refs.~\cite{Brodsky:2006uqa,%
Brodsky:2014yha,Gutsche:2013zia}. 

\section{Summary}

We have extended a soft-wall AdS/QCD model at small temperatures 
proposed in Ref.~\cite{Gutsche:2019blp} for the 
description of bosonic hadrons, 
to fermionic hadrons (baryons, pentaquarks, etc.). The approach implements 
important features of QCD at zero and low temperatures: (1) the dilaton field, 
responsible for spontaneous breaking of chiral and conformal symmetry, which 
plays an important role in the temperature dependence of hadronic properties, 
(2) the $T$-dependence, which coincides with the quark condensate dependence 
in QCD. We present analytical results for the temperature dependence of both 
the mass spectrum and form factors of fermionic hadrons. 
 
\begin{acknowledgments}

This work was funded by 
the Carl Zeiss Foundation under Project ``Kepler Center f\"ur Astro- und
Teilchenphysik: Hochsensitive Nachweistechnik zur Erforschung des
unsichtbaren Universums (Gz: 0653-2.8/581/2)'', by CONICYT (Chile) under 
Grants No. 7912010025, No. 1180232 and PIA/Basal FB0821. 

\end{acknowledgments}

\end{document}